\documentclass[twocolumn,showpacs,preprintnumbers,amsmath,amssymb,floatfix]{revtex4-1}

\usepackage{color}%option added by cyw eg \pagecolor{blue} \color{red}
\usepackage{graphicx}% Include figure files
\usepackage{dcolumn}% Align table columns on decimal point
\usepackage{bm}% bold math

\begin{document}

\title{Nuclear annihilation by antinucleons}

\author{Teck-Ghee Lee$^{1}$ and Cheuk-Yin Wong$^2$}

\affiliation{$^1$Department of Physics, Auburn University, Auburn, Alabama
  36849, USA}

\affiliation{$^2$Physics Division, Oak Ridge National Laboratory, Oak
  Ridge, Tennessee 37831, USA}

\def\bb#1{\hbox{\boldmath${#1}$}}
\def\pbar {\bar p}

\begin{abstract}
We examine the momentum dependence of $\bar p$$p$ and $\bar n$$p$
annihilation cross sections by considering the transmission through a
nuclear potential and the $\bar p p$ Coulomb interaction.  Compared to
the $\bar n p$ annihilation cross section, the $\bar p p$ annihilation
cross section is significantly enhanced by the Coulomb interaction for
projectile momenta below $p_{\rm lab} <$ 500 MeV/$c$, and the two
annihilation cross sections approach the Pomeranchuk's equality limit
[JETP Lett. 30, 423 (1956)] at $p_{\rm lab}\sim 500$ MeV/$c$. Using
these elementary cross sections as the basic input data, the extended
Glauber model is employed to evaluate the annihilation cross sections
for $\bar n$ and $\bar p$ interaction with nuclei and the results
compare well with experimental data.
\end{abstract}

\maketitle

\section{Introduction}
 In support of experiments of the FAIR (Facility for the Research with
 Antiprotons and Ions) at Darmstadt\cite{Fai09,Pan13} and the AD
 (Antiproton Decelerator) at CERN \cite{Mau99} for antimatter
 investigations, it is of interest to continue our investigation on
 the annihilation between an antinucleon with nucleons or a nucleus
 that represent an important aspect of the interaction between
 antimatter and matter.  A recent suggestion of using $\bar n A$
 annihilation to study the $n$-$\bar n$ oscillations \cite{Phi14}
 provides an additional impetus to examine the annihilation between an
 $\bar n$ and a nucleus.  In a recent work \cite{Lee14}, we extended
 the Glauber model for nucleus-nucleus collisions \cite{Gla59,Gla70,
   Won84,Won94} to study the antiproton-nucleus annihilation
 process. The extended Glauber model for the calculation of the $\bar
 p A$ annihilation cross section \cite{Lee14} consists of treating the
 nucleon-nucleus collision as a collection of binary collisions, with
 appropriate shadowing and the inclusion of initial-state and
 in-medium interactions. The basic ingredients are the elementary
 $\bar p p$ and $\bar p n$ annihilation cross sections, $\sigma_{\rm
   ann}^{\bar p p}$ and $\sigma_{\rm ann}^{\bar p n}$, together with
 initial-state Coulomb interactions and the change of the momentum of
 the antinucleon inside the nuclear medium.  The model provides an
 analytical and yet intuitive way to analyze $\bar p$-nucleus
 annihilation processes.  Qualitative features were reproduced to give
 a general map of the annihilation cross sections as a function of
 nuclear mass numbers and collision energies.

We would like to improve upon these earlier results on several
important aspects.  In our previous work, the basic $\bar p p$
annihilation cross section, $\sigma_{\rm ann}^{\bar p p}$, was
parametrized semiempirically as $1/v$, the inverse of the relative
velocity $v$, and utilized in our investigation of the stability and
the properties of matter-antimatter molecules \cite{Lee11,
  Lee08}. Such a simple dependence arises from the nuclear interaction
between $p$ and $\bar p$ in the $s$-state and gives the main feature
of the important momentum dependence of the annihilation cross
section.  Higher partial waves are also present and it is necessary to
includes them properly.  In addition to the nuclear interaction, $p$
and $\bar p$ also interact through the attractive Coulomb interaction
and $\sigma_{\rm ann}^{\bar p p}$ is expected to behave as $1/v^2$ in
the lowest energy region \cite{Wig48,Lan58}.  It is of interest to
examine the combined effects of the nuclear and Coulomb interactions
to see how the $1/v$ behavior of the $\bar p p$ annihilation cross
section is modified in the lowest energy region. A proper treatment of
the Coulomb and nuclear interactions for $\sigma_{\rm ann}^{\bar p p}$
will also lead to a better determination of $\sigma_{\rm ann}^{\bar n
  p}$, which is expected to vary as $1/v$ at the lowest energies.
Furthermore, in our earlier work in \cite{Lee14}, $\sigma_{\rm
  ann}^{\bar p n}/\sigma_{\rm ann} ^{{\bar p p}}$ was taken to be 4/5,
based on the experimental ratio $(\sigma_{\rm ann}^{\bar p
  n})_D/(\sigma_{\rm ann}^{\bar p p})_D$=$0.749\pm0.018$ for $\bar p$
at rest and $0.863\pm0.018$ for $\bar p$ in flight \cite{Kal80}, and a
model of nucleon-antinucleon annihilation by the annihilation of quark
and antiquarks of the same flavor \cite{Lee14}.  Because of the
attractive Coulomb interaction is present in $p\bar p$ annihilation
but absent in $\bar p n$ annihilation, $ \sigma_{\rm ann}^{\bar p p}$
should be greater than $\sigma_{\rm ann}^{\bar p n}$ and the ratio
$\sigma_{\rm ann}^{\bar p n}/\sigma_{\rm ann}^{\bar p p}$ should be
energy dependent.  Quark and antiquark can form a string and
subsequently fragments, independent of the flavor contents of the
quark and the antiquark.  Thus, the approximate fixed ratio of
$\sigma_{\rm ann}^{\bar p n}/\sigma_{\rm ann}^{\bar p p}$ of
Ref.\ \cite{Lee14} should be amended and its energy dependencies must
be properly taken into account.

On the theoretical side, there is the pioneering prediction of
Pomeranchuk \cite{Pom56} on the equality of the annihilation cross
section for $\bar p p$ and $\bar p n$ at high energies. One can
envisage a $q$-$\bar q$ pairing model of nucleon-antinucleon
annihilation in which the annihilation between a nucleon and an
antinucleon takes place by pairing the valence quark of any flavor
from the nucleon with any valence antiquark of any flavor from the
antinucleon, with each $q$-$\bar q$ pair forming a string that
subsequently fragments to many $\bar q q$ pairs (mesons), as in the
string fragmentation in $pp$ collisions \cite{Won94,Won15}.  At high
energies when the long range Coulomb effects become unimportant, such
a $q$-$\bar q$ pairing model will predict the equality of $
\sigma_{\rm ann}^{{\bar p n}}$=$ \sigma_{\rm ann}^{{\bar p p}}$
because there are the same numbers of nine ways to combine the $q$ and
$\bar q$ pairs to form strings in ${\bar p n}$ and ${\bar p p}$
annihilations.  An equality of $ \sigma_{\rm ann}^{{\bar p n}}$=$
\sigma_{\rm ann}^{{\bar p p}}$ at high energies will favor the
$q$-$\bar q$ pairing model and is consistent with Pomeranchuk's
prediction.  It will exclude another annihilation model, as for
example, the annihilation only by quarks of the same flavor
\cite{Lee14}.

To test Pomeranchuk's prediction and the annihilation models, we
re-examine the basic cross sections of $\sigma_{\rm ann}^{\bar p n}$
and $\sigma_{\rm ann}^{\bar p p}$ to understand their similarities as
well as their different energy dependencies.  There are no
experimental data of $\sigma_{\rm ann}^{\bar p n}$ for the collision
of a $\bar p$ projectile with an isolated neutron target in free
space.  There are however experimental $ \sigma_{\rm ann}^{{\bar n
    p}}$ annihilation cross section data using a $\bar n$ beam source
(from the $\pbar p \to \bar n n$ reaction) colliding on a liquid
hydrogen target \cite{Arm87,Ber97}, which are better suited for
nucleon-antinucleon annihilation studies than those of \cite{Kal80}
using the $\bar p$-(${}^2$H) annihilations.  As $ \sigma_{\rm
  ann}^{{\bar n p}}$=$ \sigma_{\rm ann} ^{{\bar p n}}$, we shall
therefore treat them equivalently and consider the problem of the
annihilation of $\bar p$ on $n$ to be equivalent to the problem of
$\bar n$ on the target proton $p$.

To study the Coulomb and nuclear interactions of an antinucleon on the
proton target, we shall assume for simplicity a square well potential
of a fixed depth for which analytical results can be readily obtained
\cite{Bla52}.  The theoretical results and the comparison with
experimental data allows one to draw a conclusion on Pomeranchuk's
prediction and the annihilation models. Upon the determination of the
improved basic $ \sigma_{\rm ann} ^{{\bar p p}}$ and $\sigma_{\rm ann}
^{{\bar n p}}$ annihilation cross sections, they can then be used as
the building blocks to evaluate the annihilation cross sections for
antinucleons on a nucleus.

It is worth pointing out that over the years, a large set of
experimental data in the annihilation of nucleons and nuclei by $\bar
p$ and $\bar n$ had been accumulated
\cite{Arm87,Ber97,H1,H2,H3,H4,NA,Bia11,Bia00a,Bia00b,D2,D3,He1,He2,He3,C,Ne,
  Al,Kle05} and analyzed theoretically \cite{Kle05,Lee14,Mah88,Kuz94,
  Car93,Car97,Bia00,Gal00,Bat01,Fri14,Gal02,Gal03,Gal08,Gal11,Gal11PL,Gal12,Gal13,Gal14,Gal15}.
Klempt, Batty, and Richard reviewed various phenomenological analyses
of microscopic quark dynamics and symmetry considerations in
nucleon-antinucleon annihilations.  The roles of initial- and
final-state interaction are also examined \cite{Kle05}.  A theoretical
optical potential based on the Glauber model \cite{Gla59,Gla70} has
been developed by Kuzichev, Lepikhin and Smirnitsky to investigate the
antiproton annihilation cross sections of various nuclei at the
momentum range of 0.70$-$2.50 GeV/$c$ \cite{Kuz94}.  In this range of
relatively high antiproton momenta, the Glauber model gives a good
agreement with the experimental data, with the exception of the
deviations at the momentum of 0.7 GeV/$c$ for heavy nuclei.  Batty,
Friedman, and Gal have developed a unified optical potential approach
for low-energy $\bar{p}$ interactions with proton and with various
nuclei using a density-folded optical potential \cite{Gal00,Bat01}.
They found that even though the density-folding potential reproduces
satisfactorily the $\bar{p}$ atomic level shifts and widths across the
periodic table for $A$$>$10 and the few annihilation cross sections
measured on Ne, it does not work well for He and Li.  Galoyan,
Uzshinsky, and collaborators have previously investigated cross
sections of various processes in $\bar pp$ collisions in many
different mechanisms. They have used different parametrizations of
the basic total and elastic $\bar pp$ cross sections in the Glauber
model and have successfully implemented these calculations in the
GEANT4 program for the simulation of the passage of particles through
matter in high-energy nuclear detector studies
\cite{Gal02,Gal03,Gal08,Gal11,Gal11PL,Gal12,Gal13,Gal14,Gal15}.  In
the low-momentum regime ($p_{\rm lab} < $ 1 GeV/$c$), however, many
questions remain open to provide additional motivation for the present
study. For example, how does the electrostatic Coulomb interaction
between the collision pair affects the annihilation cross section as
function of target mass $A$ and charge numbers $Z$, and the projectile
momentum in the laboratory frame $p_{\rm lab}$?  And at approximately
what momentum the contributions of the Coulomb interaction begins to
be less effective? This study attempts to address these questions.

The paper is organized as follows.  In Sec. II, we study the basic
$\bar p p$ and $\bar p n$ annihilation cross sections by considering
the effects of particles transmission through a nuclear potential
barrier, initial-state Coulomb interaction between the collision pair
and relativistic two-body kinematics.  As the results for the present
survey will not be sensitive to the fine structure of the potential
well, we shall assume a square well potential for which analytical
results for the transmission coefficients are well known.  The
experimental $\bar p p$ and $\bar n p$ annihilation cross sections can
be successfully described in terms of transmission coefficients of
various partial waves and Coulomb Gamow factors.  In Sec. III, the
basic $\bar p p$ and $\bar n p$ cross sections obtained in the
theoretical analysis is then included in the extended Glauber model to
calculate $\bar p$-nucleus collisions.  The expressions are given for
the $\bar p$-nucleus annihilation cross sections in terms of basic
$\bar p $-nucleon annihilation cross section, $\sigma_{\rm ann}^{\bar
  p-{\rm nucleon}}$.  In Sec. IV, we assess the theory by comparing
its numerical results to experimental data at both high and low
energies, Finally, we conclude the present study with some discussions
in Sec. V.

\section{Theory of $\bar p p$ and $\bar n p$ annihilation cross sections}

To analyze the $\bar p p$ and $\bar n p$ annihilation cross section at
a center-of-mass energy $E_{\rm c.m.}$, we follow Blatt and Weisskopf
\cite{Bla52} to decompose the incoming plane waves into partial waves
and we use the ingoing-wave strong absorption model to assume that a
partial wave transmitted passing through the nucleon surface $R$ will
lead to a reaction, which in our case is an annihilation. In the case
of $\bar p p$ annihilation, there is in addition the initial-state
Coulomb interactions which can be taken into account through the
Coulomb Gamow factor $G_L(k)$ \cite{gamow} (or the $K$-factor $K(\eta)$
in \cite{Won95,Won99}).  The $\bar p p$ and $\bar p n$ annihilation
cross sections for a collision with a wave number $k$=$\sqrt{2 \mu
  E_{\rm c.m.}}$ and a reduced mass $\mu$ are then given in terms of the
transmission coefficients $T_L$ and the Gamow factor $G_L$ by
\begin{eqnarray}    
\sigma_{\rm ann}(k) = \frac{\pi}{k^2}\sum\limits_{L=0}^{L_{\rm max}} (2L+1)T_L(k) G_L(k) ,
\label{1a} 
\end{eqnarray}  
where 
$G_L(k)$ is 1 for  $\bar p$$n$ annihilation.

To calculate the transmission coefficients, we consider the nucleon
and the antinucleon to interact through a nuclear interaction, which
for simplicity can be taken to be a square well $V(r)=- V_0 \Theta
(R-r)$.  The transmission coefficient is then given by Eq.\ (5.5) on
p. 360 of \cite{Bla52} as
\begin{eqnarray}
T_L = \frac{4 s_L KR}{\Delta_L^2 +(s_L + KR)^2},
\end{eqnarray}
where $K=\sqrt{k^2+2\mu V_0}$,  
\begin{eqnarray}
s_L = R \left [ \frac{ g_L(df_L/dr) - f_L (dg_L/dr)}{g_L^2 +f_L^2}\right  ]_{r=R},  \\
\Delta_L =  R \left [ \frac{ g_L(df_L/dr) + f_L (dg_L/dr)}{g_L^2 +f_L^2} \right ]_{r=R}, \\
f_L(r)=\left ( \frac{\pi k r }{2} \right )^{1/2} J_{L+1/2} (kr),\\
g_L(r)=-\left ( \frac{\pi k r }{2} \right )^{1/2} N_{L+1/2} (kr),
\end{eqnarray}
where $J_{L+1/2} (kr)$ and $N_{L+1/2} (kr)$ are Bessel and Neumann functions, respectively.   The Gamow factor 
for
 $\bar p$$p$ annihilation under the Coulomb interaction $V_c(r)=\alpha/r$  is
 \cite{Won99}
\begin{eqnarray}
G_L(k)= &&\frac{(L^2+\xi^2) [(L-1)^2+\xi^2] \cdots  (1+\xi^2)}{[L!]^2}
\nonumber\\
&&\times 
\left(\frac{2\pi \xi}{\exp\{2\pi \xi\}-1}\right),
\label{eq7}
\end{eqnarray}
where $\xi = \alpha/v$ and $\alpha$ is the fine structure constant.
Following Todorov \cite{Tod} and Eqs. (21.13a)$-$(21.13c) of Crater {\it
  et al.}  \cite{Crater}, in the center of mass coordinate system, it
is shown that the relative velocity $v$ for two equal-mass particles
with rest mass $m$ is related to their center of mass $\sqrt{s}$ and
can be expressed as \cite{Won95}
\begin{eqnarray}
v&=&\frac{(s^2 - 4 s m^2)^{1/2}}{s-2 m^2} 
\end{eqnarray}
and
\begin{eqnarray}
s&=&(a+b)^2= (a_0+b_0)^2 - (\bb a + \bb b)^2,
\label{cmmass}
\end{eqnarray}
where $a=(a_0,\bb a)$ and $b=(b_0,\bb b)$ are the four-momentum vectors of the two colliding particles with $a$ and $b$ 
represent the target and projectile, respectively.  

\section{The $\bar p p$ and $\bar n p$ annihilation cross sections}

\begin{figure}[h]
  \centering
\includegraphics[scale=0.4]{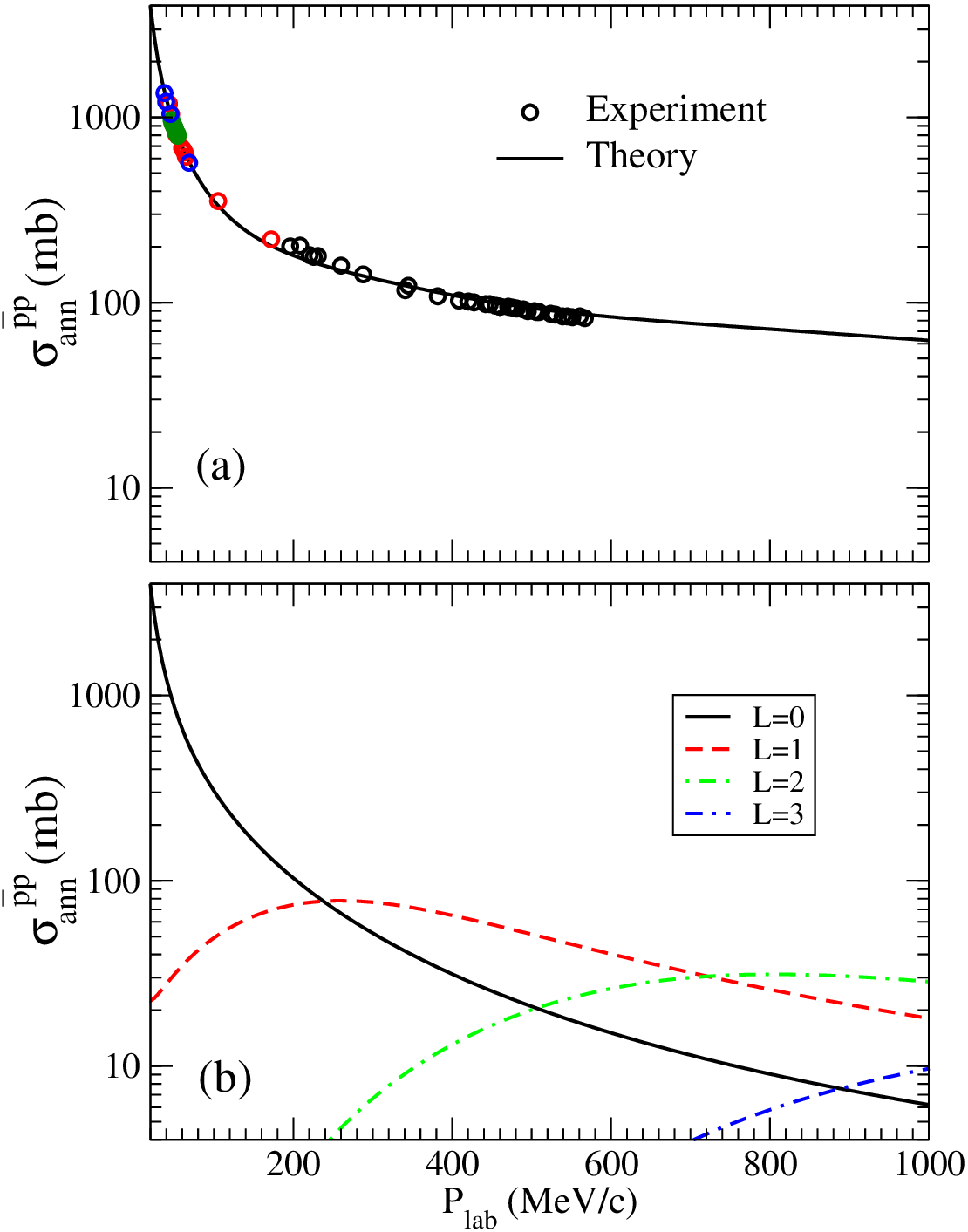}
\caption{(a) Antiproton-proton annihilation cross
  sections as a function of the antiproton momentum in the laboratory
  frame.  The solid curve represents the $\bar p p$ annihilation cross
  section of Eq.(\ref{1a}). The experimental data points are from the
  compilation of \cite{Bia11}, where the individual experimental
  sources can be found. (b) Contributions from different partial waves
  to the total annihilation cross sections.}
\end{figure}

Expression (1) shows that for the antinucleon-nucleon annihilation
cross section all necessary information is contained in the magnitudes
$T_L(k)$ and $G_L(k)$; they define the cross section completely.  To
determine $T_L(k)$ and hence the cross section, we assume the
nuclear contact radius $R = 0.97$ fm and the strong interaction
potential $V_0 = 85$ MeV. Figure 1 displays the $\sigma_{\bar p
  p}^{\rm ann}$ cross section result obtained with Eq. (\ref{1a}) as a
function of $\bar p$ incident momenta. Clearly shown in Fig.\ 1(a) is
the theoretical result fits the experimental data impressively well
over a broad energy range. The different contributions to the cross
section from $L=0-3$ partial waves as a function of energy is
demonstrated in Fig.\ 1(b).  Strong momentum dependence is observed
for all the partial waves. The $S$ wave is obviously dominated at
momentum below 240 MeV/$c$. As $p_{{\bar p}{\rm lab}}$ increases from
240 to 750 MeV/$c$, the contribution from $P$ wave becomes important. As
the incident energy increases further, i.e., above 750 MeV/$c$, the
$D$ wave begins to dominate, and so forth.

At this point, we are interested not only in the magnitude of the
cross section given in Eq.\ (\ref{1a}), but also in its behavior for
smaller values of $p_{{\bar p}{\rm lab}}$. To examine the cross
section behavior at low-energy limit, we restrict ourselves to the
case where the entrance channel wave number $k << K$ and the $S$ wave is
dominant. This simplifies the analysis and helps to elucidate the
essential points. According to Eq.\ (\ref{1a}), the annihilation cross
section is reduced to
\begin{eqnarray}
\sigma_{\bar p p} ^{\rm ann} &=& \frac{\pi}{k^2} T_0(k) G_0(k) \nonumber\\
 &=&\pi\left(\frac{4K}{k(K+k)^2}\right) \left(\frac{2\pi \xi}{\exp\{2\pi \xi\}-1}\right).
\end{eqnarray}
The first factor in the formula clearly displays the $1/v$ behavior
for $k << K$ while the parameter $\xi \to \infty$ gives
\begin{eqnarray}
  \frac{2\pi \xi}{\exp\{2\pi \xi\}-1} \to   \frac{2\pi \alpha} {v}.
\end{eqnarray}
In that event, the product of the two factors leads to $\sigma_{\bar p
  p} ^{\rm ann} \propto 1/{v^2}$ behavior at low-energy limit.  This
$1/{v^2}$ law was first pointed out by Wigner \cite{Wig48} and now its
discussions can be found in quantum mechanics textbooks \cite{Lan58,
  New82}.
        
\begin{figure}[h]
\centering
\includegraphics[scale=0.4]{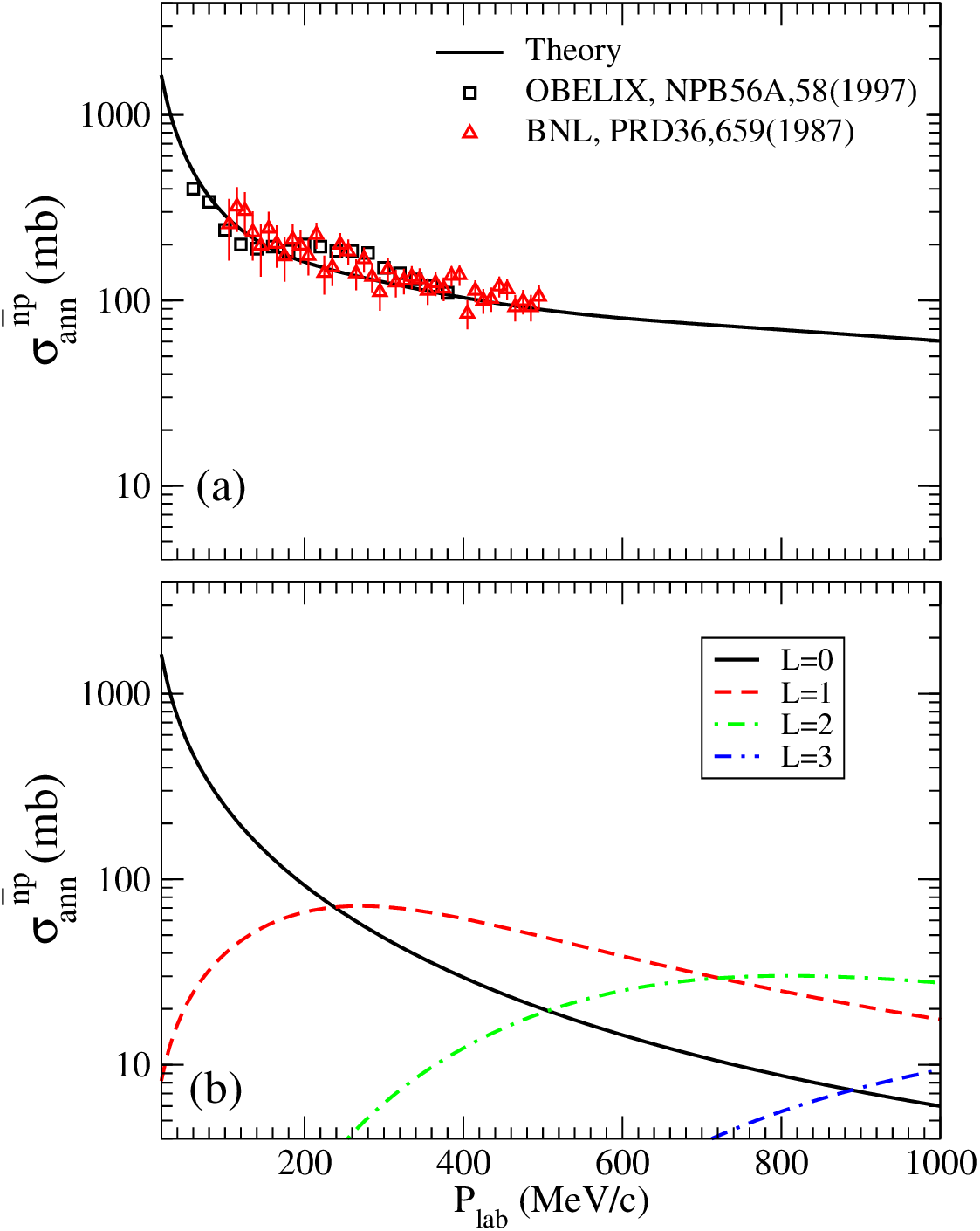}
\caption{(a) Antineutron-protron annihilation cross
  sections as a function of the antineutron momentum in the laboratory
  frame.  Solid curve: $\bar n p$ annihilation cross section using
  Eq.(\ref{1a}) with the Gamow factor equal to 1. Solid triangle:
  experimental data from the OBELIX Collaboration \cite{Ber97} and
  open circle denotes experimental data from BNL \cite{Arm87}.  (b)
  Contributions from different partial waves to the total annihilation
  cross sections.}
\end{figure}  
      
Having demonstrated that Eq.~(1) is capable of reasonably describing
the experimental $\bar p p$ annihilation cross section for a wide
momentum range, we next examine the $\bar n p$ annihilation cross
section as a function of the antineutron momentum for which
$G_L(k)=1$.  Fig.\ 2(a) shows a comparison between the theoretical and
two sets of experimental data from Brookhaven National Laboratory
(BNL) \cite{Arm87} and from the OBELIX Collaboration \cite{Ber97}.  Relative
to the $\bar p p$ measurements, the annihilation cross section data
for $\bar n p$ still remains relatively sparse to date and contain
significant degrees of uncertainties.  The two sets of data fall
within the error bars of each other.  The OBELIX data at around
$p_{\rm lab}\sim 200-300$ MeV/$c$ appears to show an enhancement
whereas the BNL data show greater fluctuation and appear to be
qualitatively consistent with the theoretical predictions.
Ultimately, we concur with Friedman's opinion \cite{Fri14} that the
broad enhancement in the experimental finding for $\bar n p$
annihilation cross sections around 200$-$300 MeV/$c$ \cite{Ber97}
remains an open question.

\begin{figure}[h]
\centering
\includegraphics[scale=0.35]{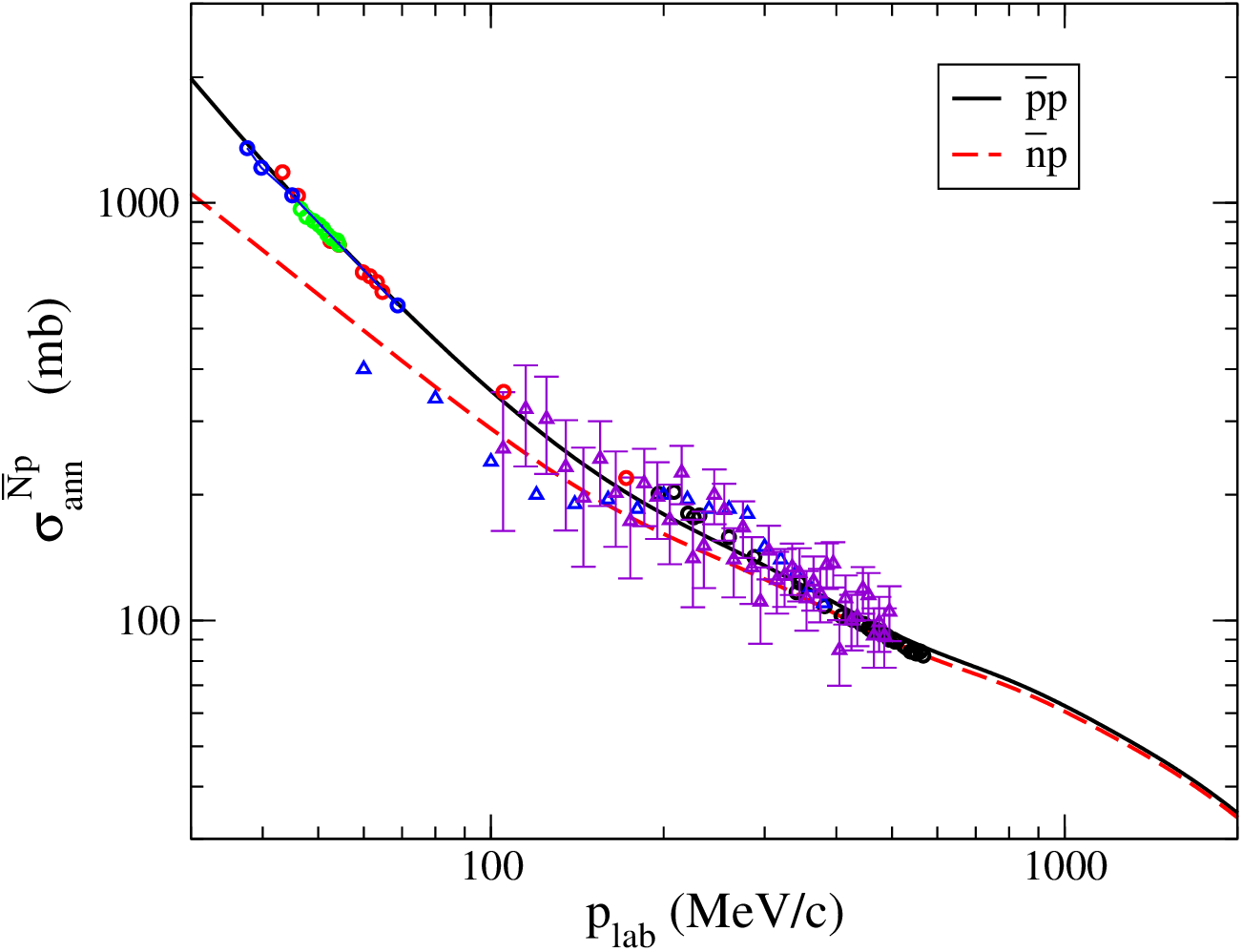}
\vspace*{0.2cm}
\caption{Comparison of $\bar pp$ and $\bar n p$ 
   annihilation cross sections as a function of the antiproton momentum 
   in the laboratory frame.}
\end{figure}
 
Figure 3 indicates the importance of the Coulomb effect by comparing
the $\bar n$ and $\bar p$ on proton annihilation as a function
energy. The theoretical data are also plotted against the available
experimental data.  At high energy limit $p_{\rm lab} >$500 MeV/$c$,
both the $\bar n p$ and $\bar p p$ curves coincide.  As $ \sigma_{\rm
  ann}^{\bar p n} $=$\sigma_{\rm ann}^{\bar n p}$, the result in
Fig.\ 3 validates the Pomeranchuk prediction \cite{Pom56} of $
\sigma_{\rm ann}^{\bar p p}$=$\sigma_{\rm ann}^{\bar p n}$ for $p_{\rm
  lab} > $ 500 MeV/$c$.

At the low-energy limit, it immediately becomes obvious that the slope
for the $\bar p p$ interaction is much steeper compared to the $\bar n
p$ one . Parametrizing the theoretical annihilation cross section in a
power law form $\sigma_{\rm ann} \propto p^x_{{\bar p}{\rm lab}}$, the
exponential value $x$ can be simply obtained via $x = \partial
ln(\sigma_{\rm ann})/\partial ln(p_{{\bar p}{\rm lab}})$. For the case
of $\bar p p$, it is found that $x$$=$$ -$1.544 in the momentum range
between 30 and 50 MeV/$c$.  Although it is not quite equal to
$x$$=$$-$2.0 as expected to be at the low-energy limit
\cite{Wig48,Lan58}, the behavior of the cross section is rightly
approaching this limit as the projectile momentum further decreases.
For the case of $\bar n p$, it is found that $x$=$-$1.080 in the
momentum range between 30 and 95 MeV/$c$. Indeed, this exponential
value is very close to the expected $x$=$-$1.0 value, a clear
indication of the $1/v$ behavior.  The cross sections of these two
cases have distinct power indices at low energies, depending on the
charged or neutral character of the interaction pair.
 
\section{The extended Glauber Model and comparison with 
experimental $\sigma_{\rm ann}^{\bar N A}$ for $A > 1$ nuclei}

The results in the last section pertain to the annihilation with $A =
1$ nucleus.  To consider $\bar p$ or $\bar n$ annihilation with
heavier $A > 1$ targets, we shall make use of our previously developed
extended Glauber model. Because the derivations of the extended
Glauber model are given in \cite{Lee14}, here we review and emphasize
only the essential formulas for describing $\pbar$ the experimental
$\bar p$-nucleus annihilation cross sections for all energies and mass
numbers.  In the extended Glauber model, we first consider the
incoming $\bar p$ travels along a linear trajectories as $\bar p$
approaches the nucleus and makes multiple collisions with the target
nucleons along its way. The target and the projectile are represented
by a density distribution function.  For the target nucleus with small
mass numbers $A < 40$, Gaussian density distribution function is
considered. On the other hand, for the target nucleus with larger mass
numbers, i.e., $A > 40$, uniform density distribution function with
sharp-cut off is considered.  The integral of the density distribution
along the $\bar p$ trajectories gives the thickness functions which,
in conjunction with the basic $\sigma_{\rm ann}^{\bar p p}$ and
$\sigma_{\rm ann}^{\bar p n}$ annihilation cross sections, determines
the probability for an $\bar p$-nucleon annihilation and consequently
the high-energy $\bar p$-nucleus annihilation cross section
\begin{eqnarray}
 \sigma_{\rm ann}^{\bar p A}(\sigma_{\rm ann}^{\bar p-{\rm nucleon}})
&=& \int d\bb b \biggl  
\{ 1 - [1-T_{\bar p p}(\bb b) \sigma_{\rm ann}^{\bar p p}]^{Z}
\nonumber \\
&&\times [1-T_{\bar p n}(\bb b) \sigma_{\rm ann}^{\bar p n}]^{N} \biggr \}
\nonumber \\
& &  \!\!\!\! \!\!\!\! \!\!\!\! \!\!\!\! \!\!\!\! \!\!\!\!=
\sum_{i=0}^Z \sideset{}{^\prime}\sum_{j = 0}^N
\left(\frac{(-1)^{1+i+j}Z!N!}{(Z-i)!(N-j)! i!j!}\right)  
\nonumber \\
&&  \!\!\!\! \!\!\!\! \!\!\!\! \!\!\!\! \!\!\!\!  \!\!\!\!\!\!\!\!\times
 (\sigma_{\rm ann}^{\bar p  p})^i 
(\sigma_{\rm ann}^{\bar p  n})^j
\int d \bb b [T_{\bar p p}(\bb b)]^i[T_{\bar p n}(\bb b)]^j,
\label{1}
\end{eqnarray} 
where $T_{\bar p p}$ and $T_{\bar p n}$ denote the thickness functions
for protons and neutrons, respectively. The argument $ \sigma_{\rm
  ann}^{\bar p -{\rm nucleon}}$ on the left-hand side stands for
$\sigma_{\rm ann}^{\bar p p}$ and $\sigma_{\rm ann}^{\bar p n}$, and
the summation $\sum'_j$ allows for all cases except when $i=j=0$.  The
$Z$ and $N$ represent the number of protons and neutrons,
respectively, in the nucleus.

\begin{figure}[h]
\centering
  \includegraphics[scale=0.35]{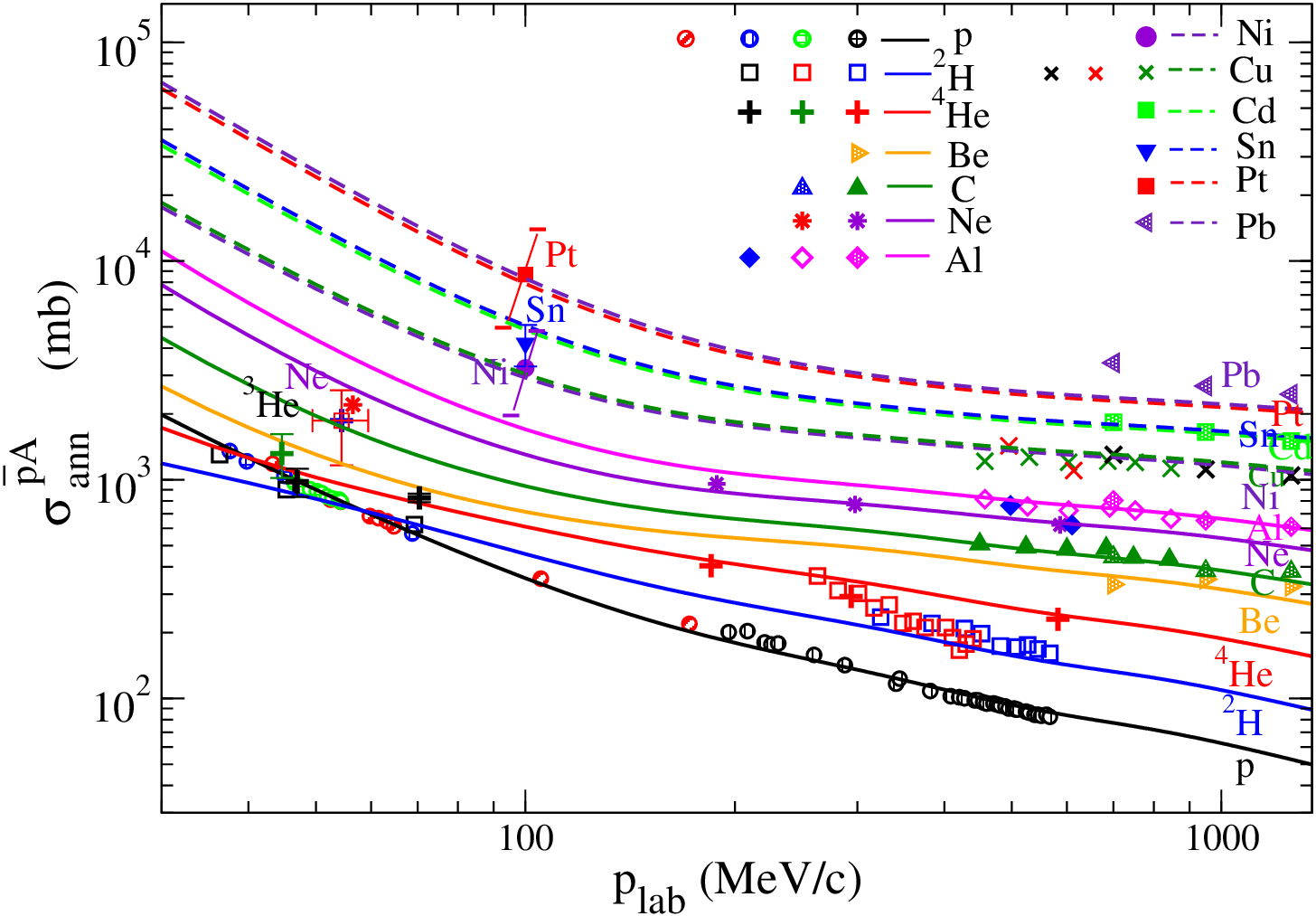}
  \vspace*{0.2cm}
\caption{$\bar{p}A $ annihilation cross sections as a
  function of the antiproton momentum in the laboratory frame. The
  experimental data are from Ref. \cite{Bia11}. The basic $\bar p p$
  curve is calculated from Eq.\ (\ref{1a}) and the rest are from the
  extended Glauber model. }
\end{figure}

To ensure Eq.(12) is also applicable for low-energy annihilation
process, we extended the high-energy Glauber model by considering the
Coulomb and nuclear interactions that are additional to those between
the incoming antiproton and an annihilated target nucleon. The
initial-state Coulomb correction resulted the modification of the
projectile trajectory from linear to curved. The strong nuclear force,
on the other hand, gives rise to the change of the antiproton momentum
in the nucleus interior. The development of the extended Glauber model
therefore resulted a compact $\bar p$-nucleus annihilation cross
section
\begin{eqnarray}
\Sigma_{\rm ann}^{\bar p A}( p_{\bar p {\rm lab}}) &=& \biggl
\{1-\frac{V_c(R_c)}{E}\biggr \} \sigma_{\rm ann}^{\bar p
  A}(\sigma_{\rm ann}^{p\bar p}(p_{\bar p {\rm lab}}'')),
\label{43}
\end{eqnarray}
where
\begin{eqnarray}
p_{\bar p {\rm lab}}''=p_{\bar p {\rm
    lab}}\sqrt{1-\frac{\left<V_c(r)\right>+\left<V_n(r)\right>}{E}}
\end{eqnarray}
represents the change of the $\bar p$ momentum inside the nucleus due
the average interior Coulomb $\left<V_c(r)\right>$ and nuclear
$\left<V_n(r)\right>$ interactions. The $\{1-V_c(R_c)/E\}$ factor on
the other hand takes into account of the initial-state Coulomb effect
that creates the path-deviation between the interaction pair from a
straight-line trajectory with $V_c(R_c)$ is the Coulomb potential
energy for $\bar p$ to be at the nuclear contact radius $R_c$ and $E$
is the center of mass kinetic energy of $\bar p$-nucleus collisions.
This analytical formula is simple. Ultimately, to evaluate the $\bar p
A$ or $\bar n A$ annihilation cross sections, one only needs to know
the fundamental $\bar p p$ and $\bar n p$ annihilation cross sections.

\begin{table} [h]
\caption{Fitting parameters.}
\begin{tabular}{cccc}
\hline
Nuclei & Gaussian &  Uniform &  ~$\langle V_n\rangle $(MeV) \\
          &~~ $r^\prime_0$(fm)~~ &  ~~$r_0$(fm) ~~&  \\
\hline
    $^2$H          &   1.20  &            &   -1.0 \\
    $^4$He &  1.20   &           &   -4.0 \\
    Be       &     1.00   &          &  -20.0 \\
    C          &   1.00  &           &  -20.0  \\
    Ne        &   1.00  &           &  -35.0  \\
    Al         &  1.00   &           &  -35.0  \\
    Ni         &           & 1.00    &  -35.0    \\
    Cu        &          &  1.00    &  -35.0  \\
    Cd        &           &  1.00    & -35.0  \\
    Sn        &           &  1.00    & -35.0  \\
    Pt         &           &  1.00    & -35.0   \\
    Pb        &           &  1.00    &  -35.0  \\
\hline
\end{tabular}
\end{table}

We consider first the $\bar p A$ annihilation cross sections. In
Fig.\ 4, the black solid curve represents the $\sigma_{\rm ann}^{\bar
  p p}$ we discussed earlier. The rest of the curves are results
obtained from the extended Glauber model with the basic $\sigma_{\rm
  ann}^{\bar p p}$ and $\sigma_{\rm ann}^{\bar p n}$ obtained in the
last section as input. Because $\sigma_{\rm ann}^{\bar p p}$ and
$\sigma_{\rm ann}^{\bar p n}$ are slightly different from those we
reported earlier \cite{Lee14}, we find it necessary to re-adjust
slightly some of the fitting parameters in the extended Glauber model
in order to reproduce the experimental results.  We use the same
functional forms and notations of the geometrical parameters as in
\cite{Lee14}.  For a light nucleus with $A < 40 $, we consider a
Gaussian thickness function with the geometrical parameter
$\beta^2=\beta_A^2+\beta_B^2+\beta_{\bar p p}^2$, where $\beta_A=r_0'
A^{1/3}$, $\beta_B=0.68$ fm, and $\beta_{\bar p p}=\sigma_{\rm
  ann}^{\bar p p}/2\pi$.  For a heavy nucleus with $A > 40$, we
consider a uniform density distribution with the sharp cut-off
thickness function and the geometrical parameters
$R_c=R_A+R_B+\sqrt{\sigma_{\rm ann}^{\bar p p}/2\pi}$, where $R_A=r_0
A^{1/3}$, and $R_B = 0.95$ fm.  The new parameters $r_0$, $r_0'$, and
nuclear potential depth $\langle V_n\rangle$ are tabulated in Table I.
Here, we also find a slightly smaller radius parameter $ r_0 = 1.00$
fm that gives a better description of the experimental data. It is worth
while to note that in the present work all the parameter values $r_0$
and $r_0'$ remain close to those used in \cite{Lee14}.

The fits to the $\bar p A$ annihilation cross sections in the present
manuscript in Fig.\ 4 are almost identical to our previous results in
Fig. 1 of Ref. \cite{Lee14}.  This indicates that the gross features
of the $\bar p A$ annihilation cross sections is insensitive to the
basic $\bar p p$ and $\bar p n$ cross sections, when the annihilation
process is properly described.  There is however only the minor
difference that with $\bar p p$ and $\bar p n $ annihilation cross
sections approaching each other at high energies, the new results
describe better the $\bar p$(${}^2$H) annihilation cross section at
around $p_{\rm lab}=400-600$ MeV/$c$.  The discrepancies of the $\bar p$(${}^2$H)  
annihilation cross section at around $p_{\rm lab}=270$ MeV/$c$
remains an unresolved theoretical and perhaps experimental problem
that needs to be rechecked.

It is illuminating to clarify why the $\bar p$Pt annihilation cross
section at $p_{\rm lab}=100$ MeV/$c$ is as large as 9000 mb,
corresponding to a black disk of annihilation with a maximum impact
parameter radius, $b_{\rm max}$, of about 17 fm, when the geometrical
touching radius, $R_{\rm Pt}+ R_{\bar p}$, is only about 8 fm.  It
should be pointed out that without the Coulomb initial-state
interaction for $\bar p$Pt annihilation, the extended Glauber model
[the second factor $\sigma_{\rm ann}^{\bar p A}$ in Eq.\ (13)] leads
to a black-nucleus result for heavy nuclei, as the $\bar p$ particle
makes multiple collisions and has many chances of annihilation with
nucleons along its path in the nucleus.  The black-nucleus cross
section obtained in the extended Glauber model is approximately $\pi
(b_{\rm max}')^2=\pi(R_{\rm Pt}+ R_{\bar p})^2 \sim \pi (8~{\rm
  fm})^2$, which is about 2000 mb.  In the presence of the Coulomb
initial-state interaction, the trajectory of a $\bar p$ at an impact
parameter $b_{\rm max}=17$ fm will be pulled down to collide with the
Pt nucleus at an impact parameter of $b_{\rm max}'=8$ fm, and the
$\bar p$ is annihilated.  The Coulomb enhancement factor
$(1-V_c(R_c)/E)$ in Eq. (13) corresponds to the ratio of $b_{\rm
  max}^2/b_{\rm max}'^2$ and is about 4$-$5, which enhances the
annihilation cross section from about 2000 mb to about 9000 mb, as
indicated in Fig.\ 4.

We consider next the $\bar nA$ annihilation cross
sections. Unfortunately, compared to the $\bar pA$ annihilation,
experiments with antineutrons are to date scarce, in particular
regarding their interaction with heavier nuclei. Nonetheless, there
are a few have been reported in literature. Figure 5 shows the
comparison of the result of extended Glauber model with the
experimental cross section for $\bar n$Fe annihilation. The data
indicate a strong dependence on the incoming $\bar n$ momentum,
similar to that of the $\bar n p$ annihilation cross section discussed
earlier. The theoretical results also appear to fit the experimental
data reasonably well, suggesting the long-range Coulomb interaction is
negligible despite the $A$ value is large.

\begin{figure}[h]
\centering
\includegraphics[scale=0.32]{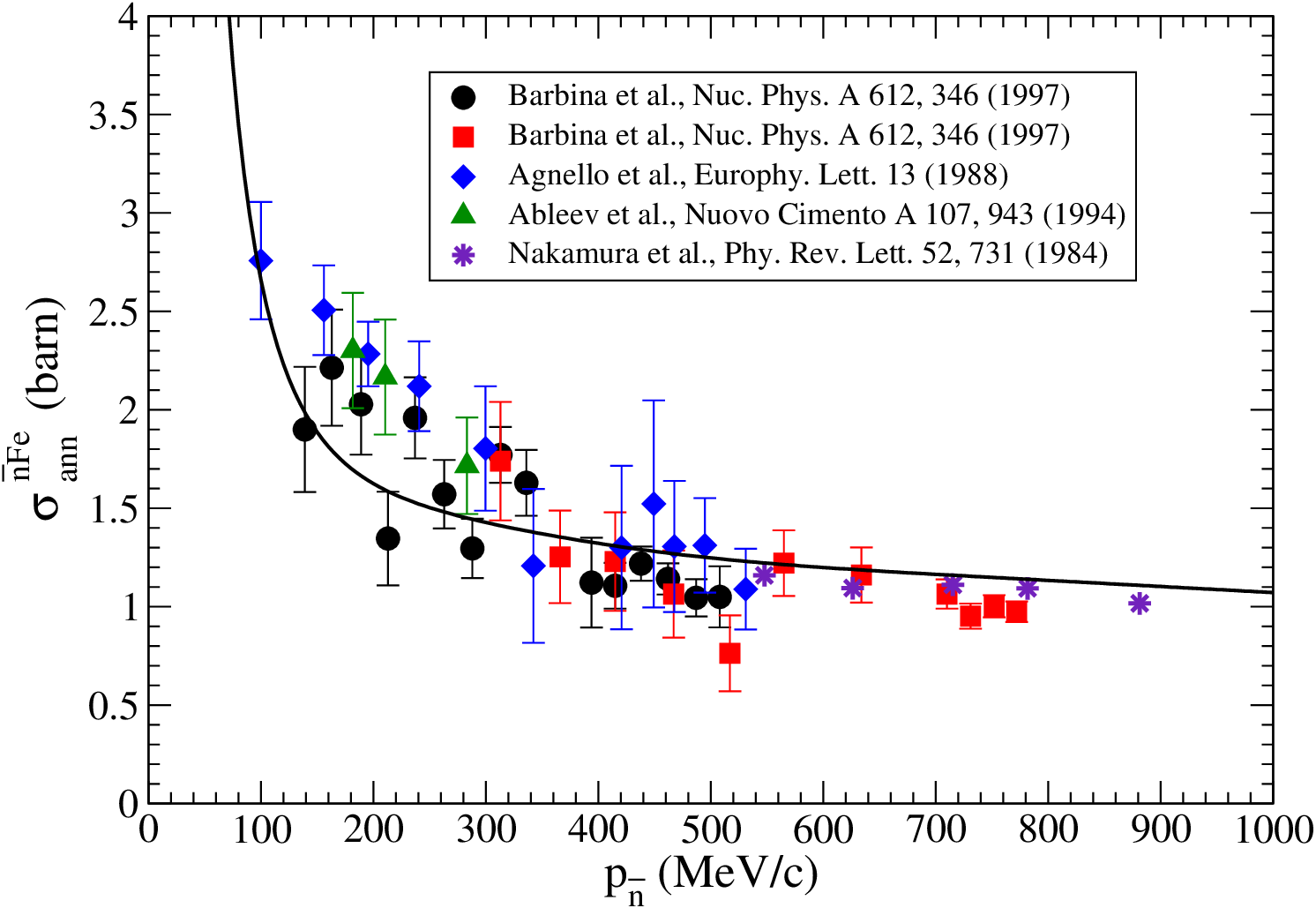}
\vspace*{0.2cm}
\caption{ $\bar{n}$Fe annihilation cross section 
as a function of the antineutron momentum in the laboratory frame. 
Comparing the result from Eq.~(1) with several sets of experimental data.}
\end{figure}

To better understand how well the present theory in describing the
$\bar nA$ annihilation, it is necessary for us to examine the
annihilation cross sections of $\bar n$ with other nuclei, namely C,
Al, Cu, Ag, Sn and Pb. Figure 6 shows the quality of agreement between
the calculations and experimental data for projectile momentum
$p_{\bar n}< 400 $MeV/$c$.

\begin{figure}[h]  
  \centering
\includegraphics[scale=0.42]{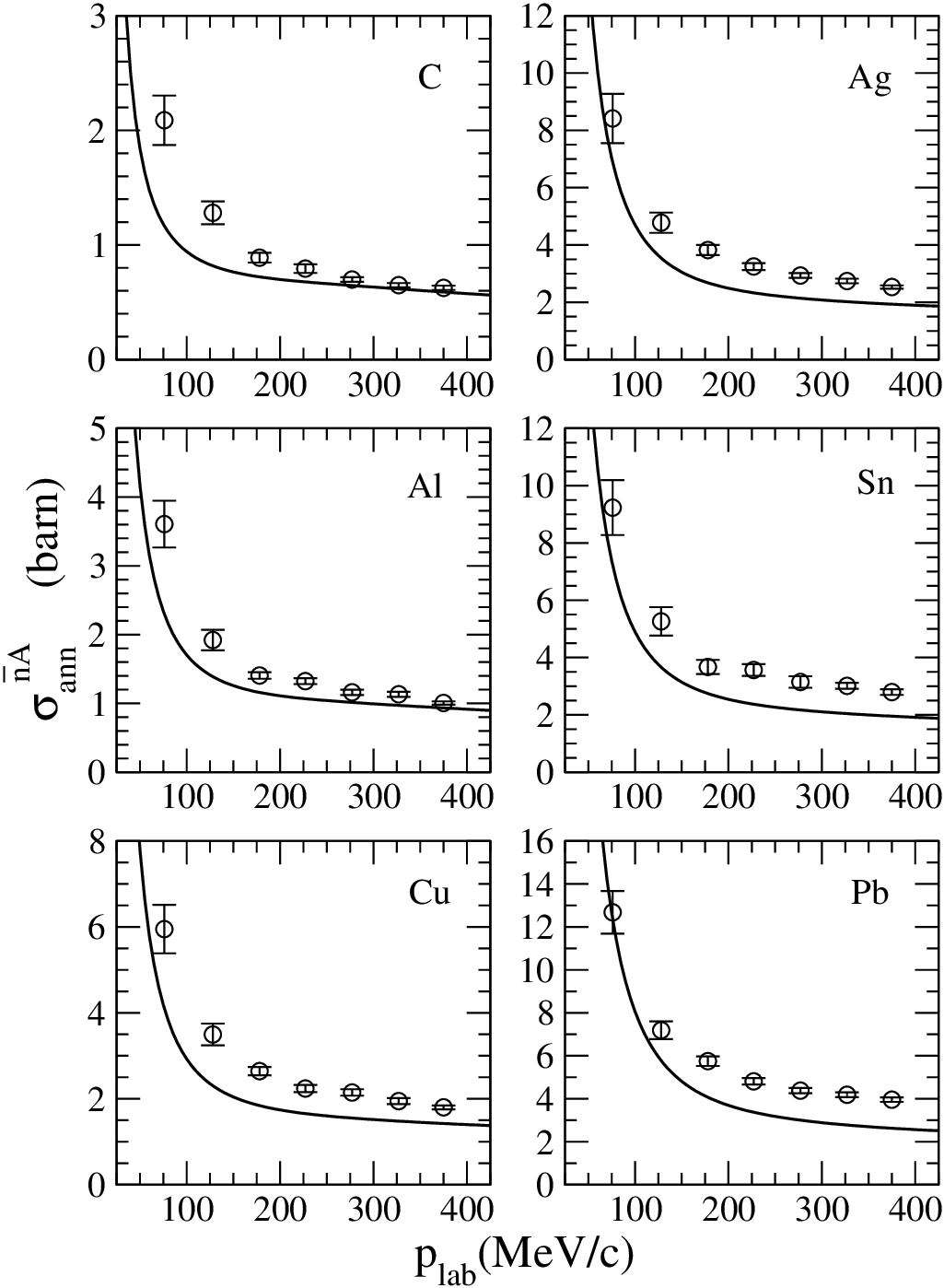}
\vspace*{0.2cm}
\caption{$\bar{n}A$ annihilation cross section
  as a function of the antineutron momentum in the laboratory frame. 
  Comparing the result from Eq.~(1) with experimental data from \cite{NA}.}
\end{figure}

In Fig. 6, we observe, for the case of C and Al targets, the agreement
between theoretical calculations and experimental data becomes poorer as
one goes down in momentum. Contrasting this with the rest of the
targets, the trend seems to go the opposite way. All said, even though
the level of the overall agreement between the theoretical and
experimental data within 20 \% is not that desirable, it is
somewhat encouraging and not to mention the extended Glauber model has
reasonably captured the main features of the annihilation cross
sections for the energy range and mass numbers
concerned. Unfortunately, at this point we cannot offer any reasonable
explanation for the origin of the discrepancy between the theory and
experiment.  But we think that both theoretical and experimental
investigations will be needed to clarify the situation.

\section{Discussion and Conclusions}

By considering the transmission through a nuclear potential and the
$\bar p p$ Coulomb interaction, the nuclear annihilation cross
sections can be properly evaluated in a simple analytical form. The
present formulation is rigorous enough and therefore amends our
earlier simple approach in which a semiempirical $1/v$ function has
been employed in order to determine the basic $ \sigma_{\rm
  ann}^{{\bar p p}}$ and $ \sigma_{\rm ann}^{{\bar p n}}$ cross
sections.  The strong absorption model formulated here decomposes the
incoming plane waves into a sum of partial waves of given orbital
angular momentum $L$ and assumes these partial waves transmit to the
nucleon surface $R$ leads to annihilation reaction.  It is shown the
cross sections for nuclear annihilation by $\bar p$ and $\bar n$ are
simple functions of the momentum of the incident particles. Across the
momenta range considered here, contrasting it to the $ \sigma_{\rm
  ann}^{{\bar n p}}$ annihilation cross section, the $ \sigma_{\rm
  ann}^{{\bar p p}}$ annihilation cross section is significantly
enhanced by the Coulomb interaction for the $p_{\rm lab}$ momenta of
the incident particle below 500 MeV/$c$. As the $p_{\rm lab}$
increases, the two annihilation cross sections become almost
identical, approaching the Pomeranchuk's equality limit at $p_{\rm
  lab}$ $\sim$ 500 MeV/$c$.  In addition, the theoretical annihilation
cross sections agree well with the experimental data. Concerning the
broad enhancement in the experimental $\bar n p$ annihilation cross
sections around 200$-$300 MeV/$c$, it is still a puzzle.
 
The equality of $\sigma_{\rm ann}^{\bar p n}$ and $\sigma_{\rm
  ann}^{\bar p p}$ at the limit of high energies predicted by
Pomeranchuk can be perceived as a $q$-$\bar q$ pairing model in which
the annihilation between a nucleon and an antinucleon takes place by
pairing the valence quark of any flavor from the nucleon to any
valence antiquark of any flavor from the antinucleon, with each
$q$-$\bar q$ pair creating a string that subsequently fragments to
many meson pairs \cite{Won94,Won15}. Such model will explain the
equality of $\sigma_{\rm ann}^{{\bar p n}}$ and $ \sigma_{\rm
  ann}^{{\bar p p}}$ when the Coulomb effects become negligible at
high energies. It overturns our naive quark model for annihilation$-$with 
annihilation takes place by pairing only the quark and antiquark
of the same flavor.

Subsequently, with the help of these elementary cross sections, the
extended Glauber model is used to evaluate the annihilation cross
sections for the $\bar p$ and $\bar n$ interaction with other nuclear
elements.  For the case of $\bar pA$ interactions, we reproduced our
previous results \cite{Lee14} and again these annihilation cross
sections are found to be in good agreement with the measurements. For
the case of $\bar n A$ interactions, predictions of the annihilation
cross section are found to be in good agreement for Fe
nuclei. However, for elements, C, Al, Cu, Ag, Sn and Pb, agreement
between the theory and experiments is found to be reasonable.

As it is now formulated, the behavior of the $\bar p A$ annihilation cross
section at low energies varies as $1/E$ arising from the Coulomb
enhancement factor, in addition to the energy dependencies of the
basic $\bar p p$ and $\bar p n$ annihilation cross sections as
described in Sec. III.  Because these basic $\bar p p$ and $\bar p
n$ annihilation cross sections increase substantially as the collision
energy decreases, the granularity nature of the individual $\bar p p$
and $\bar p n$ collisions may not play a significant role in
low-energy annihilations.  A macroscopic description of the nucleus as
a single potential without a granular structure may alternatively be a
reasonable formulation.  It will be of interest to re-examine the
antinucleon-nucleus cross section at very low energies
in a new light, by extending the potential approach as formulated in
Section III for $\bar p p$ and $\bar n p$ annihilations to $\bar p A$
and $\bar n A$ annihilations in low-energy collisions.  Future
analysis along such lines will be of great interest.

\appendix\section{Appendix} 
Following  \cite{Bla52}, the expressions for the transmission coefficient $T_L(k)$ 
for $L$ = 1, 2, and 3 partial waves
\begin{eqnarray}
T_L(k) &=& \frac{4 x X v_L }{X^2 +(2xX + x^2 v_L') v_L} 
\end{eqnarray}
with $x = kr$ and $X=KR$. The quantity $T_L(k)$ can be evaluated exactly with the functions $v_L$ and $v'_L$ given by  
\begin{eqnarray}
v_1 &=& \frac{x^2}{1+x^2},  \nonumber \\
v'_1 &=& \frac{1}{x^2}+\left( 1-\frac{1}{x^2}\right)^2,  \\
v_2 &=&  \frac{x^4}{9+3x^2+x^4},\nonumber  \\  
v'_2 &=& \left(1-\frac{6}{x^2}\right)^2 + \left(\frac{6}{x^3}-\frac{3}{x^2}\right)^2, \\ 
v_3 &=&  \frac{x^6}{225+45x^2+6x^4+x^6},  \nonumber \\ 
v'_3 &=& \left(1-\frac{21}{x^2} +\frac{45}{x^4}\right)+\left(\frac{45}{x^3}-\frac{6}{x}\right)^2.
\end{eqnarray}

Similarly, following \cite{Won99} with Eq. (76), the expressions for the Gamow 
factor
\begin{eqnarray}
G_L(\xi)= &&\frac{(L^2+\xi^2) [(L-1)^2+\xi^2] ... (1+\xi^2)}{[L!]^2} \nonumber  \\
&&\times 
\left(\frac{2\pi \xi}{\exp\{2\pi \xi\}-1}\right)  
\label{eq7}
\end{eqnarray}
for $L$ = 1, 2 and 3 partial waves can be evaluated using  
\begin{eqnarray}
G_0(\xi) &&=
\frac{2\pi \xi}{\exp\{2\pi \xi\}-1},  \\
G_1(\xi) &&= \frac{1+\frac{\alpha^2}{v^2}}{1^2}G_0(\xi),  \\
G_2(\xi) &&= \frac{[2^2+\frac{\alpha^2}{v^2}][1+\frac{\alpha^2}{v^2}]}{(2!)^2}
G_0(\xi),  \\
G_3(\xi) &&= \frac{[3^2+\frac{\alpha^2}{v^2}][2^2+\frac{\alpha^2}{v^2}][1+\frac{\alpha^2}{v^2}]}{(3!)^2}
G_0(\xi). 
\end{eqnarray}

\vspace*{0.1cm}
\centerline {\bf Acknowledgment}
\vspace*{0.2cm} 
The authors would like to thank Dr. A. Galoyan for helpful
discussions.  The research was supported in part by the Division of
Nuclear Physics, U.S. Department of Energy under Contract No.
DE-AC05-00OR22725.

\end{document}